# Surface Circular Photogalvanic Effect in Tl-Pb Monolayer Alloys on Si(111) with Giant Rashba Splitting

*Ibuki Taniuchi, Ryota Akiyama,* Rei Hobara, and Shuji Hasegawa*

Department of Physics, The University of Tokyo, Bunkyo, Tokyo 113-0033, Japan

ABSTRACT

We have found that surface superstructures made of "monolayer alloys" of Tl and Pb on Si(111), having giant Rashba effect, produce non-reciprocal spin-polarized photocurrent via circular photogalvanic effect (CPGE) by obliquely shining circularly polarized near-infrared (IR) light. CPGE is here caused by injection of in-plane spin into spin-split surface-state bands, which is observed only on Tl-Pb alloy layers, but not on single-element Tl nor Pb layers. In the Tl-Pb monolayer alloys, despite their monatomic thickness, the magnitude of CPGE is comparable to or even larger than the cases of many other spin-split thin-film materials. The data analysis has provided the relative permittivity $\varepsilon^*$ of the monolayer alloys to be ~ 1.0, which is because the monolayer exists at a transition region between the vacuum and the substrate. The present result opens the possibility that we can optically manipulate spins of electrons even on monolayer materials.

Spin degree of freedom of electrons is expected to be useful for developing devices for very low energy dissipation and quantum computing. So far, a lot of experiments of spin injection into non-magnetic materials such as semiconductors [1,2], graphene [3,4], and topological insulators [5,6] have been demonstrated using ferromagnetic electrodes and electromagnetic wave application [7,8]. Recently an alternative type of method for spin injection is rising by using spin-angular momentum of light illuminating materials having strong spin-orbit interaction (SOI). SOI lifts spin degeneracy to make voluntarily spin-split bands even for non-magnetic materials, where spin-selective optical excitations can occur with circularly polarized light (CPL) due to the conservation law of angular momentum.

Some papers already report such an optical spin injection into materials, producing spin-polarized electrical current [9,10]. Among them, circular photogalvanic effect (CPGE) attracts much attention because of its simplicity. The spin-selective optical transitions make asymmetric electron excitation in $k$-space, resulting in helicity-dependent photocurrent (HDP) flowing in a particular direction. Such non-reciprocal HDP is spin-polarized because of spin-momentum locking by strong SOI. CPGE is recently reported on many systems such as transition-metal dichalcogenides [11,12], topological insulators [13-15], and Rashba surface/interface systems [16-18]. In such two-dimensional systems, space inversion symmetry is innately broken down in the plane-normal direction, where spin-split surface bands often appear, and thus it is important to make the excitations at surfaces/interfaces dominate over those in the bulk of the substrate. For this, there are various attempts to control the excitations, such as amplifying optical responses by metamaterials [19] and band tuning with dual gates voltage [20].

In addition, it should be noted that the HDP can be induced not only by CPGE but also by photo-induced inverse spin hall effect (PISHE) [9]. Therefore, we should be careful in separating the effect incorporated in HDP. Monolayer film systems like in the present study make the interpretation simpler and can be useful for atomic-scale spintronics devices.

It is interesting to explore HDP in monolayer thin-film systems and surface superstructures because the magnitude of Rashba effect sensitively depends on the details of their atomic arrangements and species [21-23]. However, HDP on such monolayers and surface superstructures has not yet been fully studied because it had been naively believed that their thickness is too thin to produce HDP strong enough for detection, and also because *in situ* optical measurements in ultra-high vacuum (UHV) are needed to preserve the well-defined monolayer/surface structures.

In this letter, we report on strong HDP produced by CPGE with spin-selective inter-band excitations in the giant Rashba spin-split surface-state bands of surface superstructures for the first time in monolayer (Tl, Pb) alloys on Si(111) showing ($\sqrt{3}\times\sqrt{3}$) and (4×4) periodicities. We recently reported that the ($\sqrt{3}\times\sqrt{3}$) and (4×4) phases of monolayers have giant Rashba spin-split surface bands (up to an energy splitting $\Delta E \sim 250$ meV) confirmed by angle-resolved photoemission spectroscopy (ARPES), and show superconductivity at low temperatures by electrical transport measurements [24,25]. As the Si(111) substrate has a band gap of 1.1 eV, HDP induced by light, of which energy is lower than the band gap, comes from the surface-state bands of (√3×√3) and (4×4) surface superstructures only.

Since the surface Rashba systems show almost in-plane spin polarization, we need to illuminate CPL on the sample surface obliquely to inject the in-plane spin components. By reversing the polarity of CPL or the incident angle of light with respect to the surface-normal direction, the directions of HDP and spin of the flowing electrons are reversed. As a characteristic property of CPGE, when the incident beam of light is perpendicular to the surface, both charge and spin currents do not flow because the incident photons do not have in-plane spin components. In this way, CPGE can be explained by the combination of spin-split bands and selection rules at the optical transition [18].

To measure the well-defined samples in this study, we fabricated samples by molecular beam epitaxy (MBE) method, combined with *in situ* reflection-high-energy electron diffraction (RHEED) structure analysis in a custom-made UHV chamber, and sequentially measured them electrically *in situ* in the same chamber by illuminating the light without exposing the samples to air, as shown in Fig. 1(a). An example of the RHEED pattern of Si(111)-(√3×√3)-(Tl, Pb) is shown in Fig. 1(b). The Si(111) substrate was 3 mm × 10 mm × 0.5 mm in size and *n*-type moderately-doped (the resistivity $\rho$ = 1 - 5 Ω·cm at room temperature). First, Si(111)-(1×1)-Tl was prepared by depositing one monolayer (ML) Tl from a Knudsen cell onto a Si(111)-(7×7) surface held at ~ 300°C, where 1 ML = $7.8 \times 10^{14}$ cm$^{-2}$, the topmost-layer atom density of Si(111) unreconstructed surface. Then, Pb was deposited from an alumina effusion cell on Si(111)-(1×1)-Tl at room temperature. Depending on the amount of deposited Pb, the surface superstructures of (Tl, Pb) alloy layer show single phases of (√3×√3) and (4×4) periodicity at Pb coverage of 1/3 ML [25] and 2/3ML [22], respectively. Si(111)-(4×4)-(Tl, Pb) surface superstructure (Fig. 1(c)) was made by adding 1/3ML Pb atoms on Si(111)-(√3×√3)-(Tl, Pb) surface (Fig. 1(d)) after the photocurrent measurements for the (√3×√3) surface. For reference, we also made and measured single-component samples covered by 1 ML Tl only (Si(111)-(1×1)-Tl) and covered by 2 ML Pb only (Si(111)-(1×1)-Pb).

The laser beam was chopped by a light chopper, and passed through a lens, a polarizer and a quarter-wave plate (QWP), and finally introduced into the UHV chamber through an optical fused viewport (Fig. 1(a)). The light was generated by CW lasers with the wavelength $\lambda$ = 635 nm and 1550 nm, and illuminated on the center of the sample with the diameter of the laser spot ~ 1 mm. The invisible infrared (IR) laser ($\lambda$ = 1550 nm) was aligned using the guide of a visible laser ($\lambda$ = 635 nm) in the setup. The incident angle of RHEED and lasers can be changed between $\theta$ = - 60° - + 60° by rotating the sample holder as shown in Figs. 1(a) and (e). The photocurrent induced by the light illumination was measured with electrodes clamping the sample at both ends and detected by a lock-in amplifier synchronized with the chopper frequency operated at 196 Hz. The center position was defined as the midpoint of both ends where the photocurrent disappears along both *x*- and *y*-directions (Fig. 1(e)).

Figure 2 shows the light polarization dependence of the photocurrent on Si(111)-(√3×√3)-(Tl, Pb) with the laser of wavelength $\lambda$ = 635 nm (a) and (b), and 1550 nm (c) and (d), at the angle of incidence $\theta$ = - 60° (a) and (c), and $\theta$ = + 60°(b) and (d), respectively. The light polarization is changed by rotating the QWP by the angle $\alpha$; linearly polarized light ($\alpha$ = 0°, 90°, 180°), left-handed CPL ($\alpha$ = 45°), and right-handed CPL ($\alpha$ = 135°). (a) and (b) show sinusoidal changes with 90° periodicity, while (c) and (d) show changes with the periodicities of 180° as well as 90°. The photocurrent by the left-handed CPL is different from that by the right-handed CPL in (c) and (d), while they are the same in (a) and (b). The peaks and dips are reversed between (c) and (d), while they do not change between (a) and (b).

Since the photocurrent components contributed by the circularly and linearly polarized light have periodicities of 180° and 90°, respectively, the photocurrent can be expressed phenomenologically as a function of $\alpha$ by [18],

$$J = C \sin 2(\alpha + \alpha_0) + L_1 \sin 4(\alpha + \alpha_0) + L_2 \cos 4(\alpha + \alpha_0) + D. \tag{1}$$

Here, $C$ represents the component in photocurrents due to CPL, that is to say, HDP, while $L_1$ and $L_2$ represent the components due to the linearly polarized light, *i.e.*, the linear photogalvanic effect and the linear photon drag effect, respectively. $D$ is the component independent of the light polarization such as the thermoelectric effect and the photovoltaic effect. In the present study, we focus on the first term in Eq. (1) only. A phase shift $\alpha_0$ is an offset derived from the experimental setup; $\alpha_0 = 0°$ is for $\lambda = 635$ nm measurements, and $\alpha_0 = 2.7°$ and 1.6° for $\lambda = 1550$ nm measurements for Si(111)-($\sqrt{3}\times\sqrt{3}$)-(Tl, Pb) and Si(111)-(4×4)-(Tl, Pb), respectively.

To evaluate the relative magnitude of CPGE in the total polarization-dependent components including the $L_1$ and $L_2$ terms, the following equation is often used [17,26],

$$\rho_{circ} = \frac{|C|}{|C| + |L_1| + |L_2|}. \tag{2}$$

$\rho_{circ}$ for Si(111)-($\sqrt{3}\times\sqrt{3}$)-(Tl, Pb) is estimated to be 0.03 ($\lambda = 635$ nm; Fig. 2(a)) and 0.36 ($\lambda = 1550$ nm; Fig. 2(c)). This means that CPGE is negligible with $\lambda = 635$ nm while it clearly appears with $\lambda = 1550$ nm. This is reasonable because the laser with $\lambda = 635$ nm (1.95 eV) excites carriers in the Si substrate over the band gap (1.1 eV) to produce photocurrent in nA range, while the laser with $\lambda = 1550$ nm (0.80 eV) cannot excite them in the substrate, resulting in three orders of magnitude smaller photocurrent in pA range, as shown in Fig. 2. In other words, the light of $\lambda = 1550$ nm is suitable to efficiently detect CPGE from the surface superstructures of (Tl, Pb) alloys only.

The flow direction of the photocurrent by CPGE is determined by the crystal symmetry. The surface superstructures of Si(111)-($\sqrt{3}\times\sqrt{3}$)-(Tl, Pb) and (4×4)-(Tl, Pb) have $C_{3v}$ and $C_3$ symmetry, respectively. In this case, because $e_z$ element in γ-tensor [18] becomes zero (see Supplementary material), CPGE does not occur with a normal incidence of light, but the oblique incidence of light induces in-plane spin components by CPL, so that CPGE occurs. During CPGE, the electrical current flows perpendicular to the incident plane of light due to the spin-momentum locking effect and the helical spin texture [13,25].

When the light is illuminated at an angle $\theta$ in $xz$-plane as shown in Fig. 1(e), the CPGE photocurrent flows along $y$-direction, as described by [18]

$$j_y^{CPGE} = \gamma_{yx}\hat{e}_x E_0^2 P_{circ} \tag{3}$$

$$C(\theta) = A\hat{e}_x = A \cdot \frac{4\cos^2\theta \cdot \sin\theta}{(\cos\theta + \sqrt{\varepsilon^* - \sin^2\theta})(\varepsilon^*\cos\theta + \sqrt{\varepsilon^* - \sin^2\theta})}. \tag{4}$$

Here, $j$ is the photocurrent density, γ is a second-rank pseudo-tensor which contains the information of crystal symmetry of the sample and the degree of spin-split for e.g. Rashba parameter. $\hat{e}_x$ is the $x$-component of the unit vector indicating the light propagation in $xz$-plane inside the material, $E_0$ is the amplitude of electric field of the incident light, $P_{circ} = \sin 2(\alpha+\alpha_0)$ is the degree of circular polarization, and $\theta$ is the angle between the surface-normal and the direction of incident light in vacuum. $A$ is proportional to the light intensity and γ, and $\varepsilon^*$ is the relative permittivity of materials causing CPGE. The magnitude of CPGE photocurrent is described by $C(\theta)$ which corresponds to the parameter $C$ in Eq. (1). Under the normal incidence of light ($\theta = 0°$), $C(0) = 0$ means no CPGE occurs.

The experimental data of $C$ values in Eq. (1) with light of $\lambda = 1550$ nm were taken under different incident angles $\theta$ for four samples, Si(111)-($\sqrt{3}\times\sqrt{3}$)-(Tl, Pb), Si(111)-(4×4)-(Tl, Pb), Si(111)-(1×1)-Tl (1 ML), and Si(111)-(1×1)-Pb (2 ML). The results are displayed in Figs. 3 (a) and summarized in Fig.3 (b). The $\theta$ dependences of parameter $C$ in each sample were fitted by Eq. (4) as shown in Fig. 3(a) by each curve (Note: we substitute $\theta + \theta_0$ for $\theta$ in Eq. (4) for the slight misalignment of the angle as well as the fitting by Eq.(1); the resultant $\theta_0$ = - 3.0°). For both (Tl, Pb) alloy surface superstructures, the parameter $C$ is significantly enhanced with increasing $\theta$, whose tendency is consistent with that of CPGE described by Eq. (4). It should be noted here that the possibility of PISHE can be excluded; if PISHE occurs by the surface-normal component of electron spin which is injected under the normal incident of light, the parameter $C$ does not become negligibly small at $\theta$ = 0°. This is not the case in Fig. 3(a). Moreover, as shown in supplemental material, the sign of parameter $C$ is not reversed at the sample edge, which is in contrast to the case of PISHE in Ref. [9].

On the other hand, in cases of single-element metal overlayers, Tl or Pb on Si(111), the parameter $C$ is quite small at all incident angles $\theta$ in Fig. 3(a). As shown in Fig. 3 (b) their CPGE amplitudes (parameter $A$) are less than 1/8 of those of ($\sqrt{3}\times\sqrt{3}$) and (4×4)-(Tl, Pb) alloys. The reason for this small $A$ can be understood by their band structures and optical excitation processes. The surface bands of Si(111)-(1×1)-Tl have been studied in detail using spin-resolved ARPES and spin- and angle-resolved inverse-photoemission (SRIPE) measurements [27-29]; the energy splitting by the Rashba effect occurs around $\bar{\Gamma}$ point, with the energy splitting $\Delta E \sim$ 20 meV and the Rashba parameter $\alpha_R$ = 0.05 eVÅ [27], which are, however, one order of magnitude smaller than those of (Tl, Pb) alloy surface superstructures ($\alpha_R \sim$ 0.42 eVÅ for ($\sqrt{3}\times\sqrt{3}$)-(Tl, Pb) [25], and $\alpha_R \sim$ 0.27 eVÅ[22] for (4×4)-(Tl, Pb) [22]). In addition, though Si(111)-(1×1)-Tl has spin-splitting by $\Delta E \sim$ 0.6 eV above $E_F$ around $\bar{K}$ point, the spin direction is perpendicular to the surface [28,29]. This is why oblique incident light cannot induce the CPGE current. Si(111)-(1×1)-Pb also shows negligible CPGE due to the small Rashba parameter ($\alpha_R$ = 0.076 eVÅ [30]) and dense bands crossing the Fermi level $E_F$ [31]; such dense metallic bands around $E_F$ produce photocurrent irrespective of the polarization of light.

There are giant Rashba splittings in the surface bands with in-plane spin components on Si(111)-(√3×√3)-(Tl,Pb) ($\Delta E_{max}$ ~ 250 meV and $\Delta k_{max}$ ~ 0.050 Å$^{-1}$ [25]) and Si(111)-(4×4)-(Tl, Pb) ($\Delta E_{max}$ ~ 105 meV and $\Delta k_{max}$ ~ 0.047 Å$^{-1}$ [22]) near $E_F$ in the vicinity of $\bar{\Gamma}$ point, and thus electrons can be excited between them. Naturally, many optical excitations including the spin-split bands as the initial and final states can be assumed, which contribute to CPGE. The larger Rashba parameter is, the larger CPGE amplitude (parameter $A$) is usually expected, because the CPGE current includes contributions proportional to Rashba and Dresselhaus constants [32]. Although it is in general difficult to assign specific excitation paths contributing to CPGE because of many surface bands existing for possible excitations, especially on Si(111)-(4×4)-(Tl, Pb), our results qualitatively correspond to the intuitive model mentioned above.

The optical response of monolayer materials has not been investigated enough; only a few examples have been reported [11,12]. Especially the refraction of light has not been discussed for monolayer materials. Intriguingly, thanks to *in situ* measurements in our UHV systems in which we do not need any capping layer on the sample surface for protection, we have found that, from the fitting by Eq. (4), the relative permittivity $\varepsilon^*$ is almost unity for the (Tl, Pb) alloy layers (the values are shown in Fig. 3(b)). This means that, though the absorption of photons occurs at the monolayer enough to produce the HDP, the layer is extremely thin and located in a transition region between vacuum and the substrate, preventing from differentiating the relative permittivity from that of vacuum. In other words, it supports that CPGE surely comes from the surface bands of the monolayers. In previous reports for *e.g.*, topological insulators and Rashba interface systems [17,33], $\varepsilon^*$ is significantly larger than unity because of the capping layer and/or the large thickness of the samples; the refraction and transmission of light occurs at the absorptions of photons inside the materials. See the supplemental material for the detailed discussion.

In Fig. 4, let us compare our work with other reported systems in terms of the "background" band gap (ordinate), the sample thickness (abscissa), and the amplitude of CPGE $\rho_{\text{circ}}$ (color code) defined by Eq. (2). Here, the "background" band gap is defined as the bandgap inducing excitations that screen out the electrical current of CPGE. Namely, in our system, the background bandgap is that of the Si substrate. In such meaning, material systems with large background gaps are more suitable for CPGE observation. As seen in Fig. 4, first, whereas our samples are the thinnest among them, the CPGE amplitude is comparable to or relatively higher than the listed cases [11,13-16,19,26]. This benefits the atomic-layer opto-spintronics devices. Second, our monolayer samples are on the substrate having the largest band gap among listed cases, which prevents other undesired optical excitations. Although the surface superstructure is very thin so that the absorption of light had been considered too small to induce significant spin-dependent optical phenomena, we have succeeded here in showing significant CPGE.

In summary, contrary to the naive conviction, we have observed strong CPGE producing non-reciprocal photocurrent on the monolayer surface superstructures, Si(111)-($\sqrt{3}\times\sqrt{3}$)-(Tl, Pb) and (4×4)-(Tl, Pb), for the first time. This is due to spin injection to excite electrons between Rashba-type spin-split surface bands. In addition, the relative permittivity $\varepsilon^*$ is estimated to be ~ 1 from the light incident-angle dependence of CPGE. This is natural by considering that the surface superstructures are only single-atom thick and located in a transition region between vacuum and the substrate. In other words, this strongly indicates that the CPGE observed here is derived from surface superstructures. These results show that a structure with only one atomic layer can excite significant spin-polarized currents (due to spin-momentum locking) at room temperature by irradiating circular polarized light, which can then be detected electrically. This system has the potential for the next-generation monolayer devices for "surface opto-spintronics".

We thank Dimitry Gruznev and Alexander A. Saranin in Institute of Automation and Control Processes for their advice to make better samples. We also thank Shunsuke Sato for his help in the discussion and diagramming. I.T. is supported by JSPS Research Fellowships for Young Scientists and the WINGS-QSTEP program of the University of Tokyo. This research was partly supported by JSPS KAKENHI Grant Nos. 20H00342, 20H02616, and 22K18934.

FIGURES

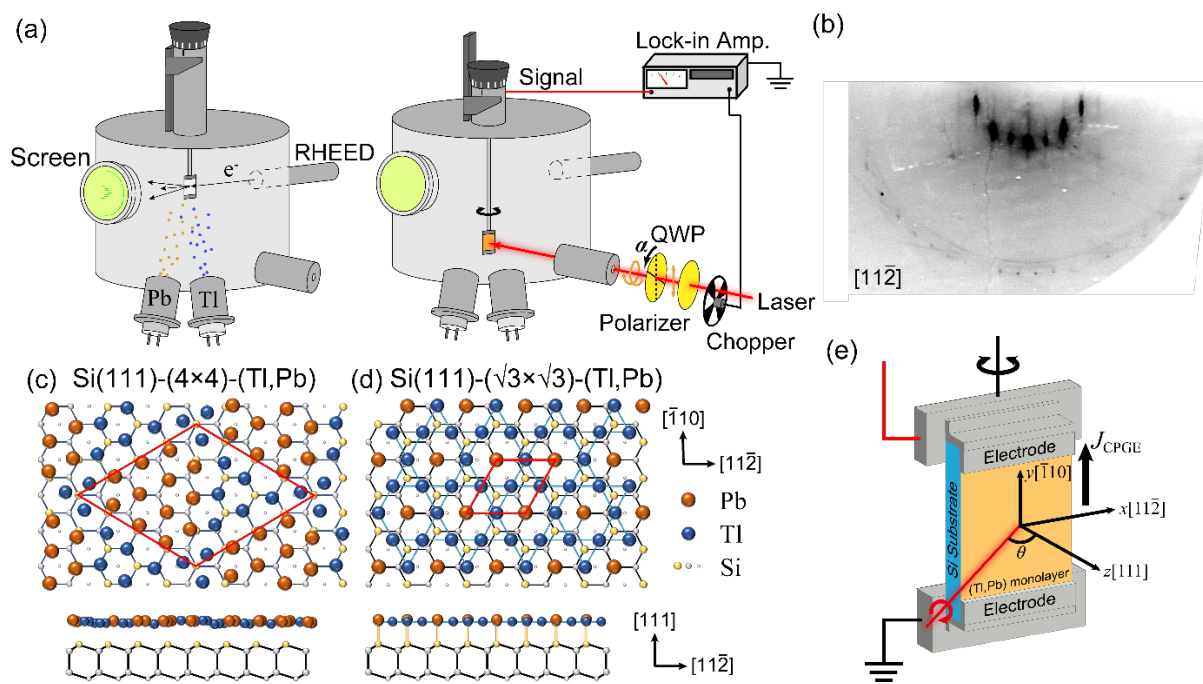

**Figure 1.** (a) Schematics of the experimental set-up during the sample growth (left) and the optical measurements (right) in the UHV chamber. (b) RHEED pattern with the incident electron beam of [11$\bar{2}$] of the Si(111) substrate just after the growth of the (√3×√3)-(Tl,Pb) monolayer on Si(111). (c,d) Top view and cross-sectional view of the atomic structures of (c) Si(111)-(4×4)-(Tl, Pb) and (d) Si(111)-(√3×√3)-(Tl, Pb) surface superstructures [22,25]. (e) Detailed figure around the sample. The photocurrent is generated by irradiating the laser on, and is detected by metal (Mo) electrodes clamping both ends of the sample. The sample can be rotated to change the angle of incidence of the laser beam.

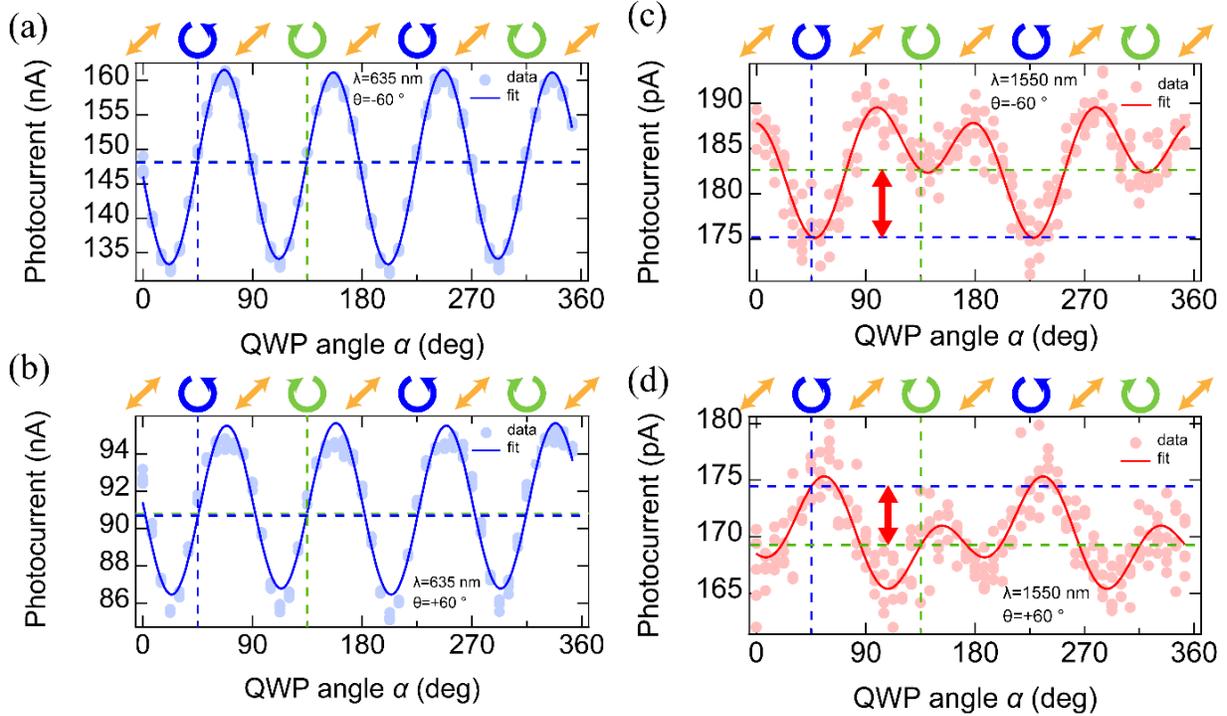

**Figure 2.** Polarization dependence of the photocurrent measured on Si(111)-($\sqrt{3}\times\sqrt{3}$)-(Tl, Pb) with (a)(b) $\lambda$ = 635 nm and (c)(d) $\lambda$ = 1550 nm, and at the angle of incidence (a)(c) $\theta$ = − 60° and (b)(d) $\theta$ = + 60°, respectively. The light was polarized by rotating the QWP with angle $\alpha$, and the polarization is shown at the top of each graph. $\alpha$ scans for five times represented by light colored dots are plotted, and deep-colored solid curves are the fitting results by Eq. (1). The green (blue) dashed line indicates the photocurrent value for the right-handed (left-handed) CPL. The difference between the two dashed lines (red arrows) corresponds to HDP, which is twice as much as the parameter $C$ in Eq. (1).

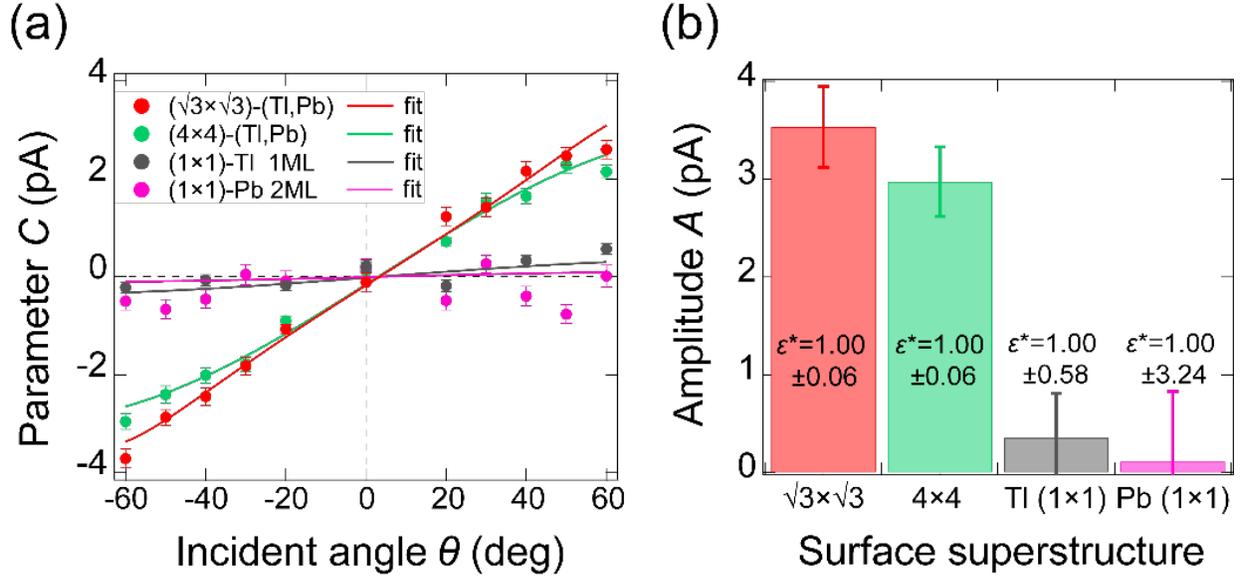

**Figure 3.** (a) Incident angle dependence of parameter $C$ in Si(111)-($\sqrt{3}\times\sqrt{3}$)-(Tl, Pb), Si(111)-(4×4)-(Tl, Pb), Si(111)-(1×1)-Tl, and Si(111)-(1×1)-Pb. The parameter $C$ was obtained by fitting the $\alpha$ dependence of the photocurrent using Eq. (1) at each incident angle $\theta$. Solid lines are fitting curves by Eq. (4) for the respective surfaces. (b) CPGE amplitude $A$ of each sample. Si(111)-($\sqrt{3}\times\sqrt{3}$)-(Tl, Pb) and Si(111)-(4×4)-(Tl, Pb) show ~ 8 times larger amplitudes than Si(111)-(1×1)-Tl and Si(111)-(1×1)-Pb. The fitting results of relative permittivity $\varepsilon^*$ are shown on center of each bar chart.

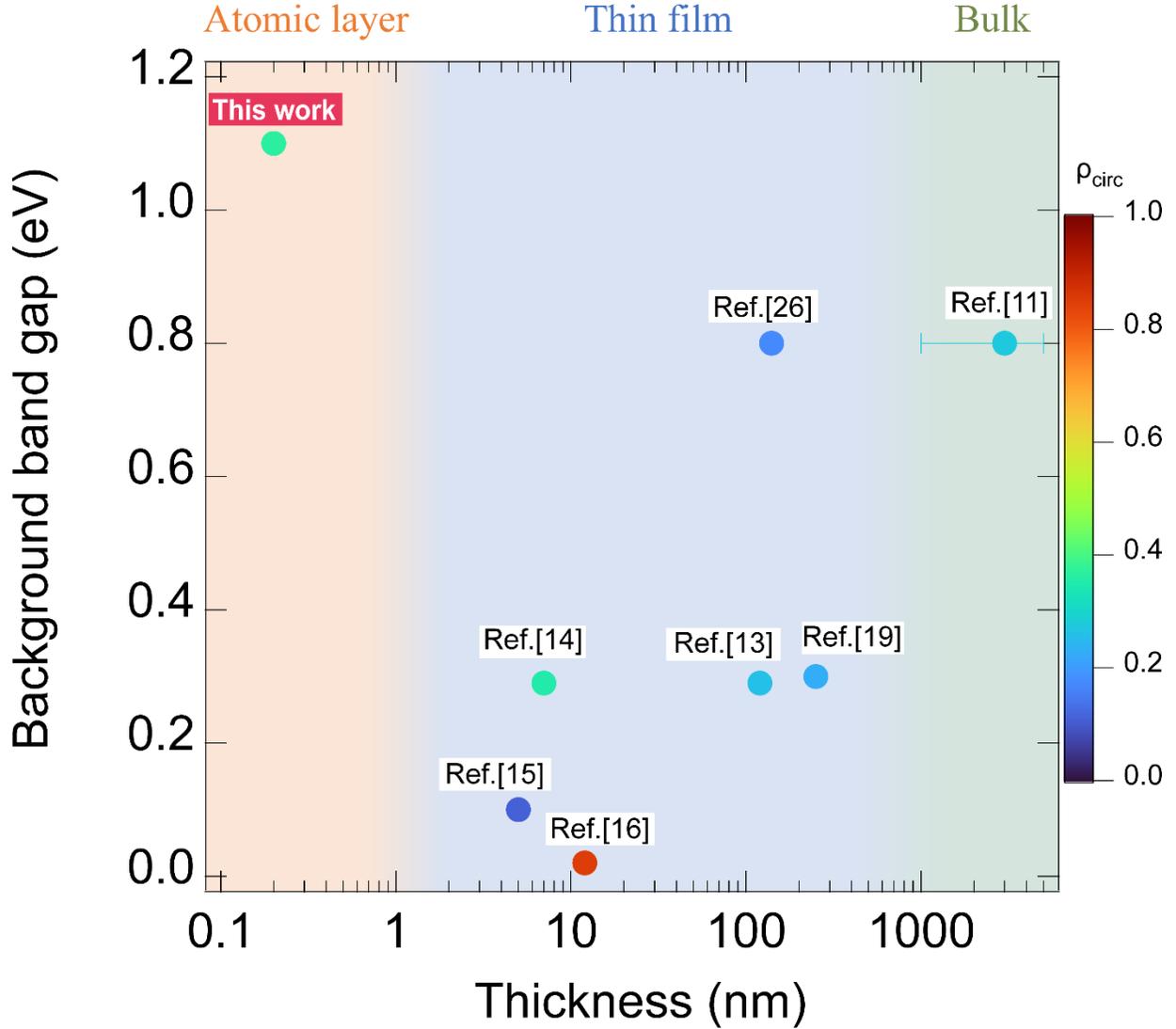

**Figure 4.** Comparison of CPGE reported in various systems. The color of each marker represents the CPGE amplitude $\rho_{\mathrm{circ}}$ defined by Eq. (2). For the thicknesses of materials, we have adopted the values available in the literature. When the values are not available, we have estimated them with error bars. For the background band gap, which is defined as the bandgap inducing excitations that screen out the electrical current of CPGE, smaller values are adopted when those of substrate and films are different in order to compare the ease of access to spin-split surface band dispersions which appear within the background band gap. The background color depicts a rough classification of samples in terms of thickness into atomic layers (orange), thin films (sky blue), and bulk materials (light green). In the case that CPGE is enhanced by some operation such as gate tuning or optical resonance, $\rho_{\mathrm{circ}}$ is determined in the bare state.

Supplementary material

# Surface Circular Photogalvanic Effect in Tl-Pb Monolayer Alloys on Si(111) with Giant Rashba Splitting

*Ibuki Taniuchi, Ryota Akiyama,\* Rei Hobara, and Shuji Hasegawa.*

Department of Physics, The University of Tokyo, Bunkyo, Tokyo 113-0033, Japan

## S1. Estimation of the PISHE component

The HDP can be induced not only by CPGE but also by PISHE. PISHE is caused optically by in-plane diffusion of spin current having out-of-plane spin or by out-of-plane diffusion of spin current having in-plane spin, and it is the combination of optical generation of spin current and inverse spin Hall effect [1-3]. Since (Tl, Pb)/Si(111) consists of monolayer Tl-Pb alloys having strong spin-orbit interaction (SOI) and the Si substrate having quite small SOI, PISHE due to spin diffusion in the depth direction is negligible. However, regarding the in-plane diffusion of out-of-plane spin component, HDP by PISHE can be observed when CPL with normal incidence is irradiated at right/left edge of the sample as we reported in the topological insulator $Bi_2Se_3$ before [1]. If PISHE is observed at a normal incidence of light, the sign of HDP should reverse at the right/left edge. Figure S1 shows the polarization dependence of the photocurrent at the left edge, center, and right edge of the sample in (Tl, Pb)/Si(111)- ($\sqrt{3}\times\sqrt{3}$) and (Tl, Pb)/Si(111)- (4×4) by normal-incidence irradiation of the laser $\lambda$ = 1550 nm. The difference between right-CPL and left-CPL, corresponding to the parameter $C$ in Eq. (1), is almost zero and the sign does not reverse between on the right and left edges. Thus, it is conceivable that PISHE is very small and the origin of HDP observed in this work is most likely due to CPGE.

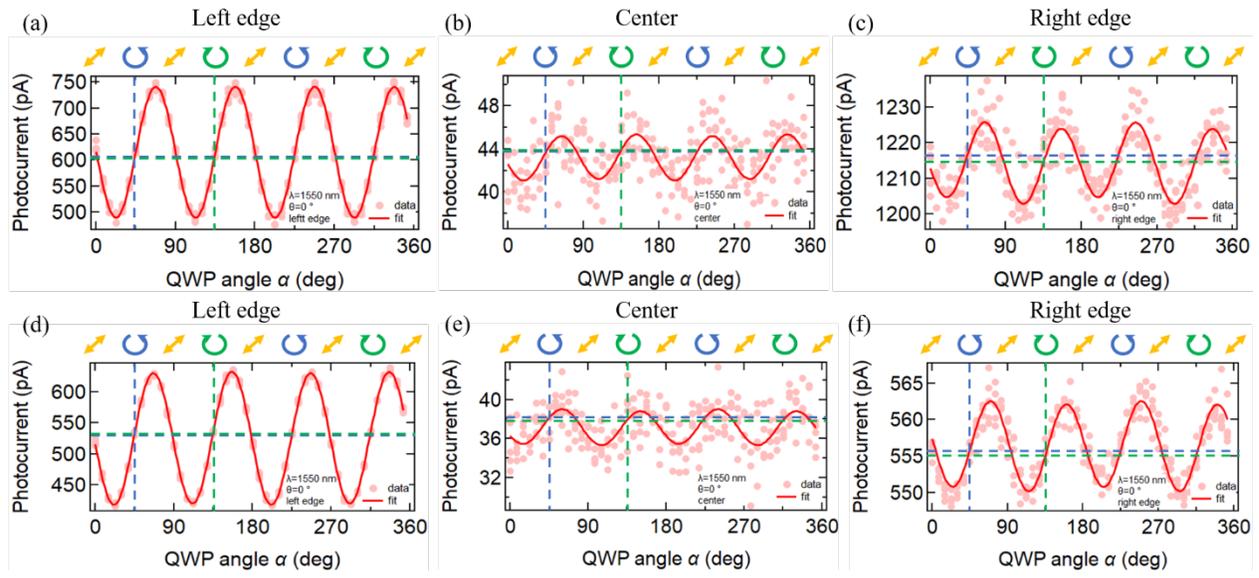

**Figure S1.** Polarization dependence of the photocurrent measured at left edge/center/right edge positions on the sample in (Tl, Pb)/Si(111)- (√3×√3) (a)(b)(c) and (Tl, Pb)/Si(111)- (4×4) (d)(e)(f) by normal-incidence irradiation of the laser λ = 1550 nm.

## S2. Surface structures observed by RHEED

Figure S2 shows RHEED patterns captured in this experiment. The crystal quality is checked *in situ* by RHEED patterns. The flux rates of Tl and Pb are calibrated using the time needed for covering the surface of Si(111)-(7×7). (a) Si(111)-(7×7), (b) Si(111)-(√3×√3)-(Tl, Pb), (c) Si(111)-(4×4)-(Tl, Pb), (d) 1 ML Si(111)-(1×1)-Tl, and (e) 2 ML Si(111)-(1×1)-Pb, respectively.

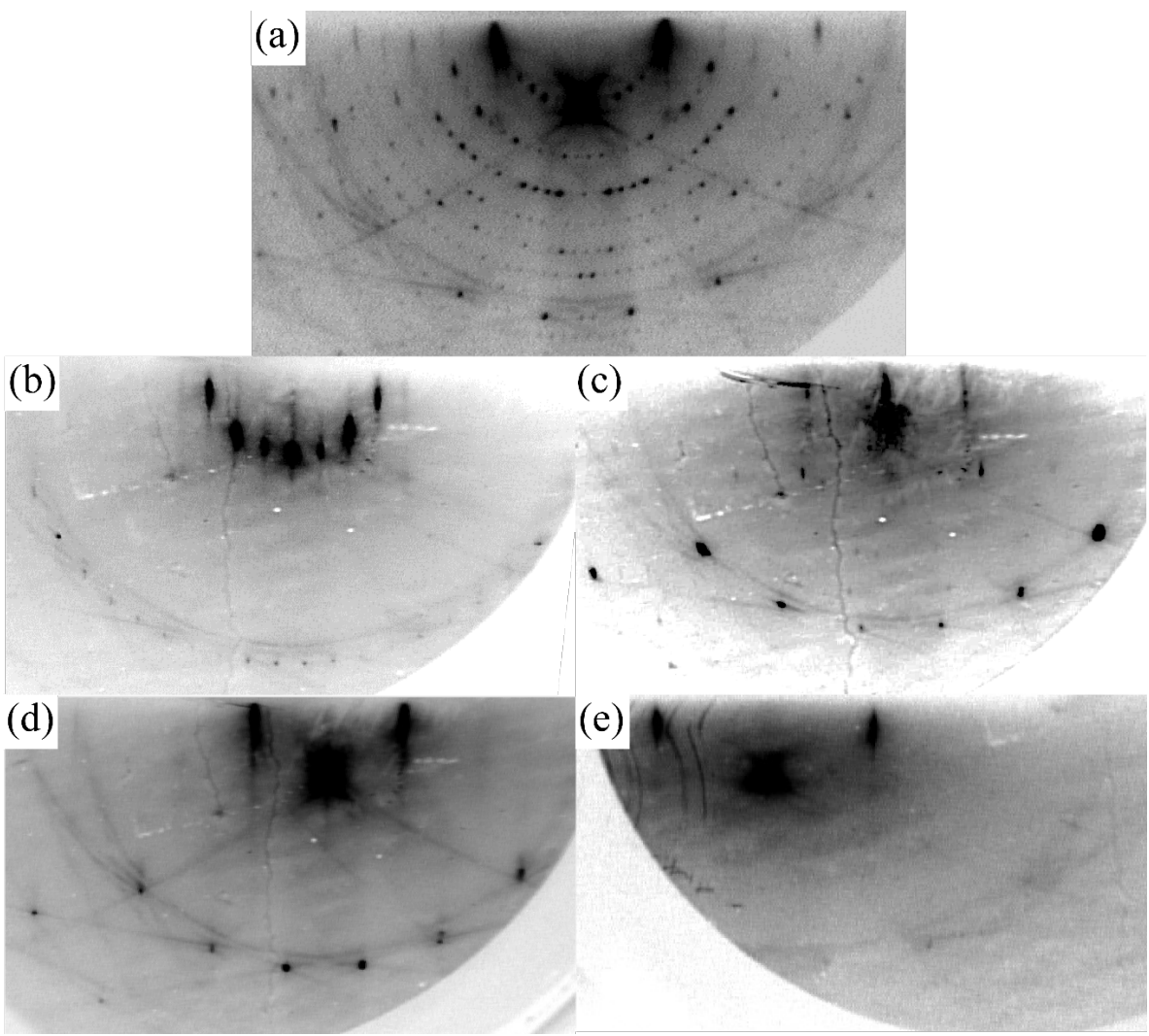

**Figure S2.** RHEED patterns observed with the electron beam of $[11\bar{2}]$ incidence on the Si(111) substrate at room temperature in this study of (a) Si(111)-(7×7) (b) Si(111)-(√3×√3)-(Tl, Pb), (c) Si(111)-(4×4)-(Tl, Pb), (d) 1 ML Si(111)-(1×1)-Tl, and (e) 2 ML Si(111)-(1×1)-Pb, respectively.

## S3. Second-rank pseudo-tensor γ for CPGE

The second-rank pseudo-tensor $\gamma$ in Eq. (3) represents the symmetry property of materials. In case a material has $T$ symmetry, $\gamma$ of the material satisfies the following equation,

$$\gamma'_{ab} = T_{ai}T_{bj}|T|\gamma_{ij}. \tag{S1}$$

Here, $\gamma'_{ab}$ is the pseudo-tensor element after the transformation, $T_{xy}$ is an element of the transforming matrix of $T$ symmetry, $|T|$ is the determinant of the transforming matrix, and $\gamma_{ij}$ is the pseudo-tensor element before the transformation. Due to Neumann's principle [4], "the symmetry elements of any physical property of a crystal must include the symmetry elements of the point group of the crystal.", for a system with $T$ symmetry, $\gamma' = \gamma$ should be satisfied. When 3-fold rotation symmetry along $z$-axis $C_{3(z)}$ and mirror symmetry to $xz$-plane $\sigma_y$ are adopted as transformation matrix $T$, the second-rank pseudo-tensor $\gamma$ is as follows;

$$C_{3(z)} = \begin{pmatrix} \cos 120° & -\sin 120° & 0 \\ \sin 120° & \cos 120° & 0 \\ 0 & 0 & 1 \end{pmatrix} : \gamma = \begin{pmatrix} \gamma_{xx} & -\gamma_{yx} & 0 \\ \gamma_{yx} & \gamma_{xx} & 0 \\ 0 & 0 & \gamma_{zz} \end{pmatrix}, \tag{S2}$$

$$\sigma_y = \begin{pmatrix} -1 & 0 & 0 \\ 0 & 1 & 0 \\ 0 & 0 & 1 \end{pmatrix} : \gamma = \begin{pmatrix} 0 & \gamma_{xy} & \gamma_{xz} \\ \gamma_{yx} & 0 & 0 \\ \gamma_{zx} & 0 & 0 \end{pmatrix}. \tag{S3}$$

In this study, $x$-direction is set along $[11\bar{2}]$ crystal orientation of the Si(111) substrate, $y$-direction is set along $[\bar{1}10]$ crystal orientation of the Si(111) substrate, and $z$-direction is set perpendicular to the Si(111) surface. Si(111)-($\sqrt{3}\times\sqrt{3}$)-(Tl, Pb) have $C_{3v}$ symmetry [5] and Si(111)-(4×4)-(Tl, Pb) have $C_3$ symmetry [6], therefore the second-rank pseudo-tensor $\gamma$ becomes the following formulas;

$$C_{3v} : \gamma = \begin{pmatrix} 0 & -\gamma_{yx} & 0 \\ \gamma_{yx} & 0 & 0 \\ 0 & 0 & 0 \end{pmatrix}, \tag{S4}$$

$$C_3 : \gamma = \begin{pmatrix} \gamma_{xx} & -\gamma_{yx} & 0 \\ \gamma_{yx} & \gamma_{xx} & 0 \\ 0 & 0 & \gamma_{zz} \end{pmatrix}. \tag{S5}$$

The CPGE current can be described on the macroscopic level by the following phenomenological expression [7];

$$j_\lambda = \sum_\mu \gamma_{\lambda\mu} \hat{e}_\mu E_0^2 P_{circ}. \tag{S6}$$

With the second-rank pseudo-tensor $\gamma$ of Eqs. (S4) and (S5) and experimental condition of $\hat{e}_y = 0$, CPGE current flows along $y$-direction and is detected with electrodes clamping the sample at the both ends, which is calculated by Eq. (3) (Note: the second-rank pseudo-tensor $\gamma$ is related to Rashba parameter [8], therefore the second-rank pseudo-tensor $\gamma$ of (Tl, Pb)/Si(111)-($\sqrt{3}\times\sqrt{3}$) and (Tl, Pb)/Si(111)-(4×4) have different values.).

## S4. Dependence of CPGE on the incident angle of light

The incident angle dependence of CPGE is represented by Eq. (4). The curve shape given by Eq. (4) depends on the relative permittivity $\varepsilon^*$ of materials. Considering the Fresnel equations and Snell's law at the boundary between material and vacuum, the unit vector of the light which propagates to $x$-direction inside of the materials is expressed by the following equation [9];

$$\hat{e}_x = t_p t_s \sin\theta', \qquad (S7)$$

where

$$t_p = \frac{2\cos\theta}{\sqrt{\varepsilon^*}\cos\theta + \cos\theta'}, \qquad (S8)$$

$$t_s = \frac{2\cos\theta}{\cos\theta + \sqrt{\varepsilon^*}\cos\theta'}, \qquad (S9)$$

$$\sin\theta' = \sin\theta/\sqrt{\varepsilon^*}. \qquad (S10)$$

Here, $t_p$ is the transmission coefficient of the p-polarized light, $t_s$ is the transmission coefficient of the s-polarized light, $\theta$ is the incident angle in the vacuum side, $\theta'$ is the angle of refraction in the material side, and $\varepsilon^*$ is the relative permittivity of the material.

Figure S3 shows curves given by Eq. (4) with various values of relative permittivity $\varepsilon^*$. The maximum and minimum positions depend on the relative permittivity $\varepsilon^*$. When $\varepsilon^* = 1$, maximum (minimum) is achieved at $\theta = 90°$ ($\theta = -90°$). When $\varepsilon^*$ increases, the maximum (minimum) position shifts to a smaller (larger) angle $\theta$, finally asymptotic to $\theta = 45°$ ($\theta = -45°$). Even if $\varepsilon^* = 2$, the maximum (minimum) position is $\theta \approx 45°$. Therefore, the parameter $C$ of (Tl, Pb)/Si(111)-($\sqrt{3}\times\sqrt{3}$) and (Tl, Pb)/Si(111)-(4×4) in Fig. 3(b) is significantly enhanced with increasing $\theta$ up to 60° because of $\varepsilon^* \sim 1.0 - 1.1$, indicating a negligible absorption of the light with the negligibly small effect of refraction.

In the refraction and transmission processes of Fresnel equations and Snell's law, the dielectric constant $\varepsilon^*$ represents how much transmission occurs and how much light is bent by refraction. In other words, the dielectric constant $\varepsilon^*$ determines attenuation coefficients $t_p$, $t_s$, and angle of refraction $\theta'$ (namely reflection coefficient). In the vacuum side $\varepsilon^* = 1.0$ whereas in the bulk side (Si substrate) $\varepsilon^* = 11.9$. Naturally, the value of $\varepsilon^*$ in the vacuum and the bulk should be smoothly connected to each other at the Si surface. Therefore, $\varepsilon^*$ of $\sim 1.0$ in the vacuum increases up to 11.9 with going deep into the bulk. The gradient of $\varepsilon^*$ in the direction of depth depends on the magnitude of $\varepsilon^*$ of the material, which in the case of Si varies with a width of $\sim 4$ nm [10-12]. Therefore, it is estimated that $\varepsilon^*$ near the surface of the monolayer superstructure is almost equal to $\varepsilon^* = 1.0$. In this way, such an effect of reduced dielectric constant near the vacuum interface is more pronounced when the film thickness is thin, which is known to be a problem, for example, when one wants to ensure capacitance in thin-film devices [13-16].

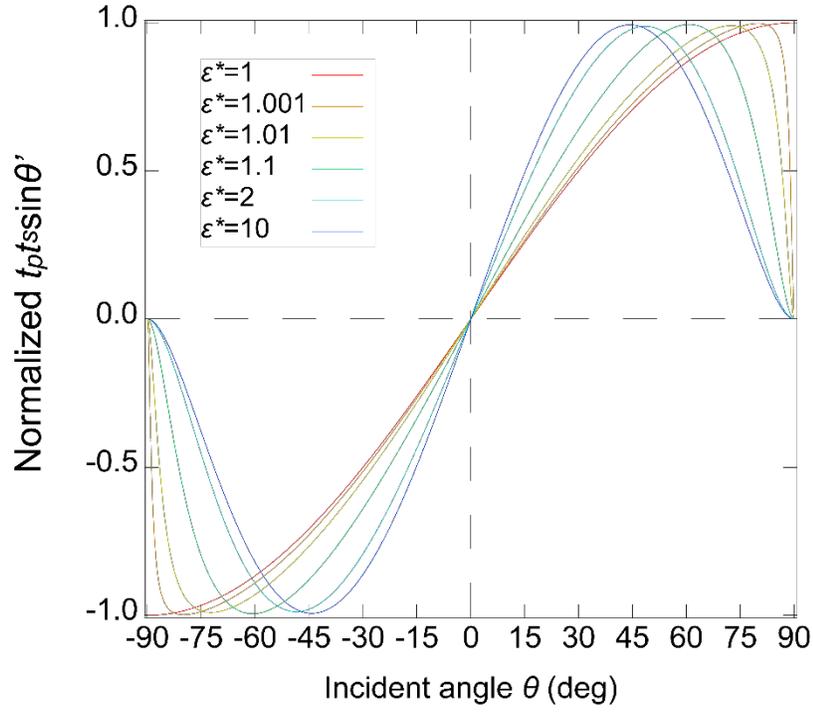

**Figure S3.** Incident angle dependences of $t_p t_s \sin\theta'$ in various values of relative permittivity $\varepsilon^*$. $t_p t_s \sin\theta'$ is normalized for comparison with the peak value; with larger $\varepsilon^*$, the peak becomes smaller.